\documentclass[prl,superscriptaddress,showpacs,twocolumn]{revtex4}

\usepackage{graphicx}

\def\figwidth{8cm}

\begin{document}

\title{Observation of Hubbard Bands in $\gamma$-Manganese}

\author{S. Biermann$^{\ast}$}


\affiliation{Institut f{\"u}r Festk{\"o}rperforschung, 
Forschungszentrum J{\"u}lich, D-52425 J{\"u}lich, Germany }

\author{A. Dallmeyer}

\affiliation{Institut f{\"u}r Festk{\"o}rperforschung, 
Forschungszentrum J{\"u}lich, D-52425 J{\"u}lich, Germany }

\author{C. Carbone}

\affiliation{Institut f{\"u}r Festk{\"o}rperforschung, 
Forschungszentrum J{\"u}lich, D-52425 J{\"u}lich, Germany }

\author{W. Eberhardt}

\affiliation{Institut f{\"u}r Festk{\"o}rperforschung, 
Forschungszentrum J{\"u}lich, D-52425 J{\"u}lich, Germany }

\author{C. Pampuch}

\affiliation{BESSY, Albert-Einstein-Stra{\ss}e 15, D-12489 Berlin, Germany}

\author{O. Rader}

\affiliation{BESSY, Albert-Einstein-Stra{\ss}e 15, D-12489 Berlin, Germany}

\author{M. I. Katsnelson}

\affiliation{University of Nijmegen, NL-6525 ED Nijmegen, The Netherlands}
\affiliation{Institute of Metal Physics, 620219 Ekaterinburg, Russia} 

\author{A. I. Lichtenstein}


\affiliation{University of Nijmegen, NL-6525 ED Nijmegen, The Netherlands}

\date{\today}

\begin{abstract}

We present angle-resolved photoemission spectra of
the $\gamma$-phase of manganese as well as a theoretical
analysis using a recently developed approach that combines
density functional and dynamical mean field methods (LDA+DMFT). 
The comparison of experimental data and theoretical predictions allows
us to identify effects of the Coulomb correlations, namely the presence
of broad and undispersive Hubbard bands in this system.
\end{abstract}

\pacs{71.20.Be,79.60.-i,71.15.Qe}
\maketitle

The electronic theory of metals is based on the concept of
quasiparticles, elementary excitations in the many-electron
system that show a one-to-one correspondence with
non-interacting electrons.
They are characterized by a
dispersion law describing the dependence of their energy on a
quasimomentum, which can be measured by angle-resolved
photoemission spectroscopy (ARPES) \cite{kevan}. 
Hubbard showed for the first time,
that strong electronic correlations can destroy this picture and
result in the formation of so-called Hubbard bands of essentially
many-body nature \cite{hubbard_III}. 
This concept is crucial for modern theories
of strongly correlated electron systems 
\cite{gkkr}. 
The formation of Hubbard bands
takes place, e.g., in many transition metal-oxide compounds,
which thus have to be viewed
as Mott insulators or doped Mott insulators 
\cite{mott}.
Transition metals represent another class of systems where  
many-body effects are important (see \cite{LKK}
and Refs. therein). However, according to common belief, they are
moderately correlated systems and normal Fermi liquids.

Electronic spectra of transition metals
have been probed intensively by angle-resolved photoemission.
Copper with its filled d-band was the first metal to be investigated
thoroughly by this technique and the results were in excellent
agreement with band structure calculations 
\cite{thiry_photoemission, knapp_cu_arpes}. The same
technique, however, showed substantial deviations when applied
to Ni and provided evidence for many-body behavior, such as the
famous $6$ eV satellite 
\cite{huefner_photoemission_ni, guillot}.
The quasiparticle damping in iron can
be as large as 30 $\%$ of the binding energy 
\cite{lda++_fe,himpsel_transition_metals}. 
Correlation effects are indeed important for metals with partially 
filled 3d bands and should be taken into account for an adequate 
description of ARPES spectra.
Nevertheless, the main part of the spectral density in Fe is related 
to usual quasiparticles, and the spectral weight of the satellite in 
Ni amounts to only 20 $\%$
\cite{himpsel_transition_metals}.

Investigations of an extended Hubbard model show that
correlation effects are strongest for half-filled d-bands \cite{zein}.
Normally the geometrical frustrations in crystals (such as in the
fcc-lattice) further enhance electronic correlations \cite{gkkr} 
so that one of the best candidates among the transition metals 
for the search of strong correlation effects is the fcc-($\gamma$) 
phase of manganese. 
It is an example of a very strongly frustrated magnetic
system; according to band-structure calculations \cite{moruzzi} the
antiferromagnetic ground state of $\gamma$-Mn lies extremely close to
the boundary of the non-magnetic phase. Moreover, an
anomalously low value of the bulk modulus \cite{guillermet} might be
considered as a first experimental hint of strong electronic
correlations.

The physical properties of bulk $\gamma$-Mn are hardly accessible in
the experiment, since the $\gamma$-phase is only stable at temperatures
between 1368 K and 1406 K, where it shows paramagnetic
behavior. Thin films of $\gamma$-Mn, however, can be stabilized by
epitaxial growth on Cu$_3$Au(100) \cite{schirmer}, which has an interatomic
spacing ($2.65 \AA$) very close to the interatomic spacing of Mn-rich
alloys ($2.60-2.68 \AA$). Schirmer et al. have shown that Cu$_3$Au(100)
supports layer-by-layer growth at room temperature up to
coverages of 20 monolayers (ML) \cite{schirmer}. A low-energy electron
diffraction (LEED) I(V) analysis revealed that the Mn films adopt
the in-plane spacing of the Cu$_3$Au(100) substrate and a
comparatively large tetragonal distortion of the fcc-lattice. For the
inner layers of a 16 ML Mn film, this distortion amounts to -6$\%$,
whereas the surface-subsurface distance is very close to the
Cu$_3$Au value.

We have used angle-resolved photoemission on the undulator
beamline TGM-5 and on the TGM-1 beamline at BESSY to probe
the electronic states in $\gamma$-Mn. The Cu$_3$Au(100) substrate was
prepared by repeated cycles of Ne$^{+}$ sputtering and annealing,
until a very good LEED pattern with sharp diffraction spots and a
low background intensity confirmed a high degree of structural
order. The base pressure of 
$2\times10^{-10}$ mbar rose to $7\times10^{-10}$ mbar
as Mn was deposited by electron beam evaporation. To avoid
interdiffusion of Cu and Au, the onset of which was determined to
be above room temperature \cite{schirmer} we used to keep the sample at
room temperature during the Mn deposition and the photoemission
measurements. The high quality of our Mn samples was routinely
verified by means of LEED and Auger spectroscopy.

Angle-resolved photoemission measures the electron spectral
density $A({\bf k},E)$ as a function of the quasimomentum $\bf k$ and the
energy $E$ multiplied by the Fermi distribution function $f(E)$ 
\cite{kevan}. For
a given electron emission angle corresponding to a given $\bf k$ the
spectral density usually has a well-defined maximum as a function
of $E$ that determines the quasiparticle dispersion $E({\bf k})$ for the
occupied part of the electronic bands. The experimental data (Fig.~1a) 
obtained for $\gamma$-Mn at a photon energy of $34$ eV and for
different electron emission angles, however, are characterized by two
striking features. These are a weakly dispersive quasiparticle band
near the Fermi level $E_F$ and a broad and almost {\bf k}-independent
maximum at approximately $2.5$ eV below $E_F$%
\footnote{The $2.5$ eV feature is insensitive to 
oxygen adsoption and therefore
not related to surface-localized states.}. 
These structures lack
a significant dispersion also in spectra taken in normal electron
emission for photon energies from $14$ to $70$ eV (Fig.~2). This
behavior clearly distinguishes $\gamma$-Mn from other transition
metals investigated with angle-resolved photoemission, which
are strongly dispersive \cite{rader}. 

\begin{figure}
\centerline{\includegraphics[width=9cm,height=10cm]{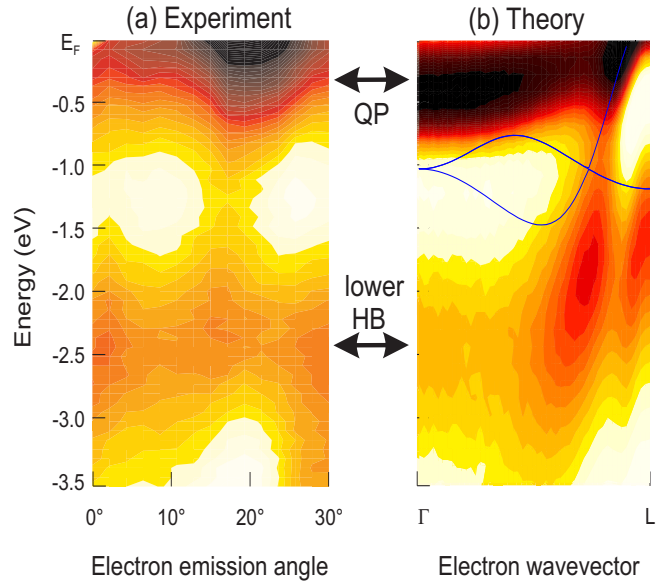}}
\caption{\label{fig1}
Experimental photoemission spectra taken at a photon energy
of $34$ eV for different electron emission angles (a) in comparison
with the spectral function $A({\bf k},\omega)$ of $\gamma$-Mn as 
calculated within the LDA+DMFT approach (b). The {$\bf k$}-values 
corresponding to the experimental data vary approximately
between the $\Gamma$ and the $L$ point in the Brillouin zone,
binding energies are measured with respect to the Fermi energy.
Colors from white, yellow, orange, red, brown to black denote
increasing intensities. 
The blue lines in (b) give the LDA band structure.} 
\end{figure}

\begin{figure}
\centerline{\includegraphics[width=12cm,height=13cm]{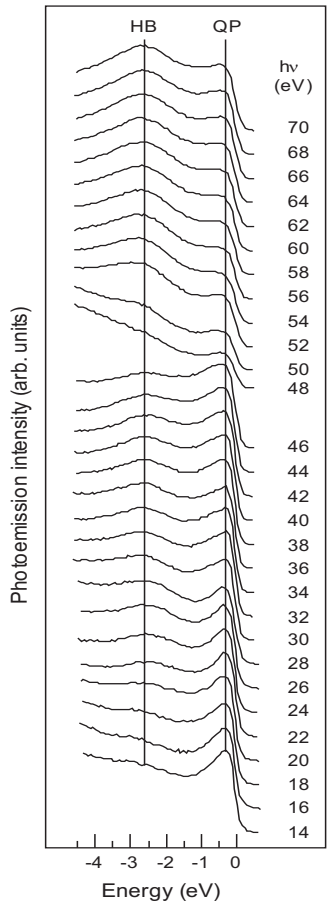}}
\caption{\label{fig2}
ARPES spectra of a $17$-monolayer
$\gamma$-Mn film taken in normal emission at photon energies
of $14$ to $70 eV$. The lack of dispersion distinguishes $\gamma$-Mn
from other transition metals. Note that the spectral changes
from $48 eV$ to $52 eV$ are due to resonant transitions between
$3p$ and $3d$ states.
} 
\end{figure}

These data cannot be understood in the framework of a standard
quasiparticle picture, since first-principles calculations of the band
structure for different magnetic phases of $\gamma$-Mn show an energy
dispersion of more than $1.5$ eV \cite{crockford}. 
Instead, the overall shape of the
experimental spectra is very close to that of the Hubbard model on
the metallic side of the Mott transition with a quasiparticle band
near the Fermi level and a broad Hubbard band below $E_F$ \cite{georges82}. 

To test this hypothesis we have carried out first-principle (LDA+DMFT) 
calculations \cite{ani1,lda++} of the electronic
structure of $\gamma$-Mn that include correlation effects in a local but
fully dynamical approximation for the electron self-energy. 
In this approach a realistic description of the delocalized s and p electrons
within the local density approximation (LDA) is supplemented by a term
describing the partially localized nature of the $d$-states. 
The Hamiltonian thus reads
\begin{eqnarray}\label{LDAUham}
H &=& H^{LDA}
+\frac 12\sum_{imm^{\prime }\sigma }U_{mm^{\prime }}^in_{im\sigma
}n_{im^{\prime }-\sigma }  \nonumber \\
&+&\frac 12\sum_{im\neq m^{\prime }\sigma }(U_{mm^{\prime }}^i-J_{mm^{\prime
}}^i)n_{im\sigma }n_{im^{\prime }\sigma },
\end{eqnarray}
where $a_{im\sigma}^{+}$ [$a_{im\sigma }$]
creates [destroys] an electron with spin
$\sigma$ in state $m$ at site $i$ and 
$n_{im\sigma }=a_{im\sigma}^{+} a_{im\sigma }$
is the corresponding number operator.
$U_{mm^{\prime }}$ and $J_{mm^{\prime}}$ are the direct
and exchange term of the screened Coulomb interaction:
$U_{mm^{\prime }} =\langle mm^{\prime }|V_{scr}({\bf r-r}^{\prime
})|mm^{\prime }\rangle$
and 
$J_{mm^{\prime }} =\langle mm^{\prime }|V_{scr}({\bf r-r}^{\prime
})|m^{\prime }m \rangle$, which can be expressed in terms of the
average Coulomb and exchange interaction parameters $U$ and $J$,
the values of which are known ($U\sim 3$eV, $J\sim 0.9$ eV)%
\footnote{We also performed calculations for other values of $U$ ($4$
  eV, $5$ eV). Higher $U$ values slightly shift the general features but do not
  lead to a qualitatively different behavior.}
\footnote{Moreover, we use the fact that for transition metals
  the ratio of the Slater integrals $F_2/F_4$ is to a good approximation
  constant and equals $0.0625$. Cf. V.~I.~Anisimov, F.~Aryasetiawan,
  A.~I.~Lichtenstein, J.~Phys.: Condens.~Matter, {\bf 9} 767 (1997)} 
.
We use an LDA-LMTO \cite{andersen1} effective Hamiltonian $H^{LDA}$,
corrected for double counting of the Coulomb energy of the
$d$ states in the usual way \cite{lda++}.
In Eq.(\ref{LDAUham}) the sums run over the $3d$
states only, whereas in the LDA Hamiltonian, $4s$, $3d$ and $4p$ states
are included.
The Coulomb interaction term is treated within the
dynamical mean-field theory (DMFT) approach, which
is the most efficient local approach: it
reduces an original many-body lattice problem to the solution of
an effective quantum impurity model in a self-consistent electron
bath \cite{gkkr}. 
This multiband impurity problem has been solved by a
numerically exact quantum Monte Carlo scheme 
based on the
algorithm of Hirsch and Fye \cite{hirsch}.
Using 64 or 128 slices in imaginary time allows to reliably access
temperatures down to $T \sim 500K$. This is still higher than
in the experiment; however, test calculations show that 
higher temperatures only result in a slight smoothening
of the spectra, therefore justifying the comparison with
the experimental data at lower temperature.
Typically, about $10^5$ QMC sweeps and 10 to 15 DMFT iterations
are sufficient to reach convergence.
The main quantities that we calculate and analyze are: a) the
local Green's function, b) the {\bf $k$}-resolved local Green's function
\begin{eqnarray}\label{integralgleichung}
\hat{G}({\bf k}, \tau) = 
\frac{1}{\beta} \sum_n e^{-i \omega_n \tau}
\left(i \omega_n + \mu - \hat{H}^{LDA}({\bf k}) - 
\hat{\Sigma}(i \omega) \right)^{-1}
\end{eqnarray}
where $\omega_n$ are the Matsubara frequencies corresponding to
the inverse temperature $\beta$.
Inversion of the spectral representations of these functions 
by means of 
a Maximum Entropy scheme \cite{jarrell}
yields the density of states (DOS) $\rho(\omega)$
and the spectral function $A({\bf k},\omega)$.
To our knowledge these calculations are the first ones that determine
the ${\bf k}$-dependence 
of the spectral density for a material with $d$-states from 
LDA+DMFT with a realistic five-band Coulomb vertex.

\begin{figure}
\centerline{\includegraphics[width=\figwidth,height=10cm]{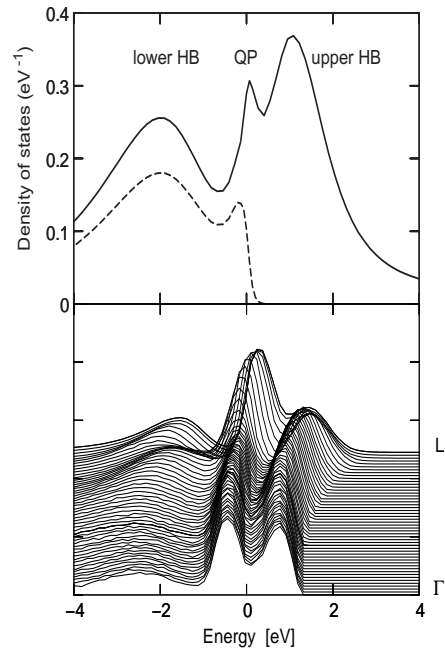}}
\caption{\label{fig3}
Upper panel: 
density of states of $\gamma$-Mn as calculated within LDA+DMFT.
The ``three-peak structure'' with the two broad Hubbard bands (HB)
and a narrow quasiparticle (QP) Kondo resonance at the Fermi level
(solid line) is typical of strongly correlated systems. The 
calculated photoemission spectrum (dashed line), i.e. the
density of states multiplied with the Fermi function and broadened
with the experimental resolution, shows reasonable agreement with
the experimental spectra Fig.~1a and Fig.~2.
Lower panel: ${\bf k}$-resolved density of states [arbitrary units]
as calculated within LDA+DMFT.
The different curves correspond to {\bf $k$}-points between
the $\Gamma$ and the $L$-point.
}
\end{figure}

The results are shown in Fig.~1b.
For a given k-point there are two energy regions that carry the main
part of the spectral weight: one narrow quasi-particle (QP) 
feature near the Fermi
level and a very broad Hubbard band at about $-2.5$ eV. 
Given the facts that (i) the experiments are done at a somewhat lower
temperature than the calculations, that (ii) we do not take into
account matrix elements for interpreting the photoemission data
and that (iii) using the Maximum Entropy scheme for determining the
spectral function, a quantity {\it not} directly measured within
the Quantum Monte Carlo simulations, introduces a further approximation,
the theoretical spectral function agrees reasonably well with the
experimental data (Fig.~1a). 
Also plotted in Fig.~1b are the Kohn-Sham eigenvalues taken from
the LDA calculation. The absence of LDA bands in the energy 
region carrying most of the spectral weight in this {$\bf k$}-space
direction is striking and underlines the necessity of a proper
many body treatment as done in LDA+DMFT. Note that assuming
antiferromagnetic order (of the type detailed below) would slightly
shift the LDA bands. However, the antiferromagnetic LDA band
structure displays a dispersion of more than $2$ eV and could thus
not explain the undispersive photoemission feature.

The calculated ({$\bf k$}-integrated and {$\bf k$}-resolved) 
density of states curves
(Fig.~3) demonstrate a characteristic "three-peak structure", with
two broad Hubbard bands and a narrow quasiparticle Kondo
resonance at the Fermi level which is typical of strongly correlated
electron systems \cite{gkkr}. The quasiparticle peak at the Fermi level and
the lower Hubbard band are seen in the present ARPES spectra; 
in k-unresolved (BIS) measurements \cite{speier}
a broad peak has been observed at $1.4$ eV. To identify this peak
with the upper Hubbard band (located at $1.2$ eV in our calculations)
one should prove the dispersionless nature of this peak. We have
checked that all these incoherent features do not depend on the
directions in {$\bf k$}-space used in our calculations. For the above
reasons we believe that $\gamma$-Mn belongs to the class of strongly
correlated materials and that the ARPES
data can be considered as the first observation of Hubbard bands
in a transition metal. 
As discussed above, correlation effects are 
indeed observed in other transition metals,
e.g. the Ni satellite or some broadening of the quasi-particle
bands in Fe. Still, even if the mechanism leading to these features
is of the same origin, their spectral weight is not comparable to
the weight of the Hubbard bands in $\gamma$-manganese.

The energy scale associated with the correlation effects 
that lead to the formation of the Hubbard bands ($\sim U$) 
is much larger than that of the magnetic interactions.
Therefore the observed effects are not very sensitive to long-range
magnetic order. 
We have carried out the electronic structure
calculations for both the paramagnetic and the antiferromagnetic
structure with wave vector Q=($\pi$,0,0), which is typical of
$\gamma$-Mn-based alloys \cite{fishman}. 
The magnetic ordering changes the electron spectrum
little in comparison with the nonmagnetic case. However, in
comparison with the results of standard band theory \cite{moruzzi}, the
correlation effects stabilize the antiferromagnetic structure leading
to a magnetic moment of about 2.9 $\mu_B$. 

According to the present results, $\gamma$-Mn can be considered a
unique case of a strongly correlated transition metal. An even
larger correlation would transform the system to a Mott insulator
where every atomic multiplet forms its own narrow but dispersive
Hubbard band \cite{hubbard_III,mott}.
On the other hand, in most metals correlations are
small enough for the quasiparticles to be
well-defined in the whole energy region and usual band theory
gives a reasonable description of the energy dispersion.
Note that the correlation strength and bandwidth
have almost the same magnitude for all $3d$ metals.
$\gamma$-Mn is
probably an exceptional case among the transition elements due to
the half-filled d-band and geometric frustrations in the
fcc-structure.

In conclusion, our ARPES data for the $\gamma$-phase
of manganese and their theoretical analysis 
by means of LDA+DMFT, an approach that accounts
not only for band structure effects on the LDA level but also
allows for a full description of local effects of strong 
Coulomb correlations, provide evidence for the formation of
Hubbard bands in metallic manganese. This is a qualitatively
new aspect in the physics of transition metals.

\begin{acknowledgments}
Acknowledgements: 
This research has been supported by
a grant of supercomputing time at
NIC J{\"u}lich and by Netherlands Organization 
for Scientific Research (NWO project 047-008-16).

$\ast$ Present address: 
LPS, CNRS-UMR 8502, UPS B\^at. 510, 91405 Orsay, France

\end{acknowledgments}

\end{document}